\documentclass[twocolumn]{cinc}
\usepackage{graphicx}
\usepackage{comment}
\newcommand{\heading}[1]{\textbf{#1}}

\usepackage{xcolor}

\begin{document}
\bibliographystyle{cinc}

\title{Estimation of Cardiac and Non-cardiac Diagnosis \\ from Electrocardiogram Features}

\author {Juan Miguel Lopez Alcaraz$^{1}$, Nils Strodthoff$^{1}$ \\
\ \\ 
 $^1$ Carl von Ossietzky Universität Oldenburg, Oldenburg, Germany.}

\maketitle

\begin{abstract}
Ensuring timely and accurate diagnosis of medical conditions is paramount for effective patient care. Electrocardiogram (ECG) signals are fundamental for evaluating a patient's cardiac health and are readily available. Despite this, little attention has been given to the remarkable potential of ECG data in detecting non-cardiac conditions. In our study, we used publicly available datasets (MIMIC-IV-ECG-ICD and ECG-VIEW II) to investigate the feasibility of inferring general diagnostic conditions from ECG features. To this end, we trained a tree-based model (XGBoost) based on ECG features and basic demographic features to estimate a wide range of diagnoses, encompassing both cardiac and non-cardiac conditions. Our results demonstrate the reliability of estimating 23 cardiac as well as 21 non-cardiac conditions above 0.7 AUROC in a statistically significant manner across a wide range of physiological categories. Our findings underscore the predictive potential of ECG data in identifying well-known cardiac conditions. However, even more striking, this research represents a pioneering effort in systematically expanding the scope of ECG-based diagnosis to conditions not traditionally associated with the cardiac system. 

\end{abstract}

\section{Introduction}

Fast diagnoses in medical settings are vital for prompt and effective treatment, significantly impacting patient outcomes. In emergencies, rapid identification of conditions such as myocardial infarctions, strokes, and infections enables healthcare providers to initiate immediate interventions, reducing morbidity and mortality rates \cite{bouzid2021rapid}. Similarly, in non-emergency but crucial scenarios, early detection of diseases allows for timely management and intervention, preventing severe complications and improving long-term health outcomes \cite{10.1093/rheumatology/kev289}.

The electrocardiogram (ECG) is a crucial tool for evaluating a patient’s heart health. Currently, ECGs are primarily interpreted manually, with minimal help from rule-based devices that have notable limitations \cite{schlapfer2017computer}. The advent of machine learning has generated excitement for AI-enhanced ECG interpretation, transforming diagnostic approaches. Many studies highlight the accuracy of machine learning in detecting various cardiac conditions and some non-cardiac conditions. However, these studies often rely on closed-source datasets and lack external validation, limiting their generalizability \cite{kulkarni2023machine,ahn2022development}.

In this work, we present a comprehensive diagnostic analysis of both traditional and non-traditional cardiac conditions, along with a variety of non-cardiac conditions, using ECG features combined with basic patient demographics like age and gender.

\section{Background}

\heading{Cardiac conditions from ECG} Although significant, the estimation of cardiac conditions has been the most conventional machine learning application with the use of ECG data, while most of these works use ECG raw waveforms \cite{ribeiro2020automatic,kashou2020comprehensive,strodthoff2020deep,strodthoff2024prospects,alcaraz2024mdsedmultimodaldecisionsupport}, some other works provide also valuable applications based on ECG features instead of raw waveforms such as \cite{rahul2021improved} which uses ECG features instead of waveforms to predict arrhythmia and \cite{sun2009predicting} for atrial fibrillation. 

\heading{Non-cardiac conditions from ECG} The estimation of non-cardiac diagnoses from ECG data has been somehow limited, where most of the approaches are too narrowed in scope as they aim to predict single conditions such as \cite{kulkarni2023machine} diabetes and pre-diabetes, \cite{ahn2022development} cirrhosis, and \cite{kwon2020artificial} pulmonary hypertension. However, recently, two main works present a wide range set of non-cardiac conditions estimation with a focus on the emergency department setting, \cite{strodthoff2024prospects} using ECG raw waveforms, while \cite{alcaraz2024mdsedmultimodaldecisionsupport} also includes also clinical features in a multimodal setting. 

\heading{Novel applications from ECG} While most of the works in the literature focus on the estimation of cardiac conditions with ECG data, recently, new paradigms have emerged such as the estimation of laboratory values abnormalities \cite{alcaraz2024cardiolab}, as well as patient deterioration in the emergency department \cite{alcaraz2024mdsedmultimodaldecisionsupport}.

\section{Methods}

\heading{Datasets} In this work, we considered two main datasets, the first is the MIMIC-IV-ECG \cite{MIMICIVECG2023,johnson2023mimic,goldberger2000physiobank} and the second is ECG-VIEW-II \cite{kim2017ecg}. To match both datasets in terms of features we engineered MIMIC-IV-ECG features to align with ECG-VIEW ones. The final set of features are sex, age in years, RR interval in milliseconds (ms), PR segment in ms, QRS complex in ms, QT interval in ms, corrected QT (QTc) interval in ms, P wave axis in degrees, QRS axis in degrees, T wave axis in degrees. For this work, we aim to investigate a wide range of diagnoses in an end-to-end manner represented by the International Classification of Diseases Clinical
Modification (ICD10-CM) codes. For MIMIC-IV-ECG we follow \cite{strodthoff2024prospects} stratified splits based on patient, age, and gender bins with the ratios of 18:1:1 for train, validation, and test sets, whereas for ECG-VIEW-II we applied a similar stratification approach ourselves. We train our models on MIMIC-IV-ECG while we use ECG-VIEW-II to externally validate our approach.

\heading{Models and performance evaluation}
We train individual extreme gradient boosting (XGBoost) tree models for each ICD10-CM code in a binary classification context. The performance is assessed individually using the area under the receiver operating curves (AUROC). To evaluate the statistical uncertainty due to the finite size and specific composition of the test set, we use empirical bootstrap with 1000 iterations on the test set and report 95\% confidence intervals.

\section{Results}

\begin{table*}[htbp]
\caption{\label{tab:cardiac} Performance results for cardiac conditions, presented by ICD10-CM code, description, internal and external performance by AUROC with 95\% confidence intervals. }
\vspace{4 mm}
\centering
\begin{tabular}{llcc} \hline\hline
\textbf{Code}  & \textbf{Description} & \textbf{Int. AUROC (95\% CI)} & \textbf{Ext. AUROC (95\% CI)} \\ \hline
I481          & Persistent atrial fibrillation                        & 0.877 (0.874, 0.878)              & 0.769 (0.768, 0.770)              \\
I44           & AV and left bundle-branch block                       & 0.873 (0.873, 0.874)              & 0.874 (0.874, 0.875)              \\
I50           & Heart failure                                        & 0.835 (0.835, 0.835)              & 0.828 (0.828, 0.828)              \\
I420          & Dilated cardiomyopathy                               & 0.822 (0.818, 0.819)              & 0.770 (0.768, 0.769)              \\
I495          & Sick sinus syndrome                                 & 0.810 (0.810, 0.811)              & 0.813 (0.810, 0.812)              \\
I350          & Nonrheumatic aortic stenosis                         & 0.805 (0.804, 0.806)              & 0.830 (0.828, 0.830)              \\
I07           & Rheumatic tricuspid valve diseases                   & 0.804 (0.803, 0.805)              & 0.832 (0.832, 0.832)              \\
I08           & Multiple valve diseases                             & 0.802 (0.800, 0.800)              & 0.808 (0.808, 0.808)              \\
I255          & Ischemic cardiomyopathy                              & 0.788 (0.787, 0.792)              & 0.852 (0.852, 0.852)              \\
I45           & Other conduction disorders                           & 0.784 (0.783, 0.785)              & 0.720 (0.717, 0.717)              \\
I850          & Esophageal varices                                   & 0.774 (0.773, 0.774)              & 0.791 (0.787, 0.788)              \\
I210          & STEMI myocardial infarction, anterior wall           & 0.763 (0.762, 0.766)              & 0.763 (0.762, 0.768)              \\
I864          & Gastric varices                                      & 0.757 (0.745, 0.749)              & 0.766 (0.762, 0.763)              \\
I110          & Hypertensive heart disease with heart failure         & 0.753 (0.752, 0.753)              & 0.846 (0.845, 0.846)              \\
I120          & Hypertensive chronic kidney disease (Stage 5)       & 0.754 (0.751, 0.754)              & 0.768 (0.767, 0.769)              \\
I27           & Other pulmonary heart diseases                       & 0.737 (0.735, 0.736)              & 0.767 (0.765, 0.766)              \\
I714          & Abdominal aortic aneurysm, without rupture            & 0.737 (0.734, 0.737)              & 0.764 (0.764, 0.764)              \\
I36           & Nonrheumatic tricuspid valve disorders               & 0.734 (0.730, 0.737)              & 0.733 (0.732, 0.737)              \\
I652          & Occlusion and stenosis of carotid artery             & 0.730 (0.729, 0.730)              & 0.790 (0.790, 0.791)              \\
I46           & Cardiac arrest                                       & 0.728 (0.726, 0.726)              & 0.731 (0.731, 0.732)              \\
I05           & Rheumatic mitral valve diseases                      & 0.726 (0.725, 0.729)              & 0.743 (0.741, 0.742)              \\
I340          & Nonrheumatic mitral valve insufficiency              & 0.721 (0.718, 0.719)              & 0.735 (0.733, 0.735)              \\
I24           & Other acute ischemic heart diseases                  & 0.710 (0.709, 0.710)              & 0.735 (0.733, 0.735)              \\ \hline\hline
\end{tabular}
\label{table}
\end{table*}

\heading{Cardiac conditions} Table \ref{tab:cardiac} presents the predictive performance results for various cardiac conditions. Overall, the model demonstrates strong generalization across a diverse array of cardiac conditions, with internal AUROC values ranging from 0.70 to 0.87.

\begin{table*}[htbp]
\caption{\label{tab:non-cardiac} Performance results for selected non-cardiac conditions, presented by ICD10-CM code, description, internal and external performance by AUROC with 95\% confidence intervals.}
\vspace{4 mm}
\centering
\begin{tabular}{llcc} \hline\hline
\textbf{Code}  & \textbf{Description} & \textbf{Int. AUROC (95\% CI)} & \textbf{Ext. AUROC (95\% CI)} \\ \hline
E282 & Polycystic ovarian syndrome & 0.952 (0.951, 0.952) & 0.890 (0.890, 0.892) \\
M329 & Systemic lupus erythematosus & 0.916 (0.914, 0.915) & 0.711 (0.712, 0.712) \\
S721 & Pertrochanteric fracture & 0.862 (0.858, 0.862) & 0.869 (0.868, 0.869) \\
M100 & Idiopathic gout & 0.846 (0.844, 0.849) & 0.722 (0.721, 0.722) \\
H353 & Macular degeneration & 0.839 (0.838, 0.841) & 0.735 (0.736, 0.736) \\
M97 & Periprosthetic joint fracture & 0.834 (0.848, 0.860) & 0.710 (0.709, 0.710) \\
K4090 & Inguinal hernia, non-recurrent & 0.815 (0.814, 0.814) & 0.747 (0.743, 0.744) \\
M797 & Fibromyalgia & 0.793 (0.794, 0.795) & 0.715 (0.711, 0.715) \\
R000 & Tachycardia, unspecified & 0.786 (0.785, 0.786) & 0.725 (0.731, 0.731) \\
R57 & Shock & 0.775 (0.777, 0.778) & 0.724 (0.720, 0.722) \\
E1121 & Type 2 diabetes with nephropathy & 0.750 (0.752, 0.752) & 0.776 (0.771, 0.772) \\
E1122 & Type 2 diabetes with chronic kidney disease & 0.747 (0.749, 0.751) & 0.735 (0.738, 0.739) \\
E29 & Testicular dysfunction & 0.747 (0.754, 0.758) & 0.773 (0.773, 0.773) \\
N18 & Chronic kidney disease & 0.743 (0.742, 0.743) & 0.740 (0.740, 0.740) \\
J441 & COPD with acute exacerbation & 0.743 (0.746, 0.747) & 0.891 (0.890, 0.890) \\
A419 & Sepsis, unspecified & 0.742 (0.741, 0.741) & 0.798 (0.797, 0.798) \\
N10 & Acute pyelonephritis & 0.735 (0.737, 0.742) & 0.737 (0.735, 0.737) \\
J86 & Pyothorax & 0.715 (0.715, 0.715) & 0.736 (0.733, 0.733) \\
J91 & Pleural effusion & 0.714 (0.711, 0.711) & 0.761 (0.762, 0.762) \\
R13 & Aphagia and dysphagia & 0.708 (0.708, 0.708) & 0.726 (0.728, 0.729) \\
R001 & Bradycardia, unspecified & 0.708 (0.711, 0.711) & 0.783 (0.786, 0.787) \\ \hline\hline
\end{tabular}
\label{table}
\end{table*}

\heading{Non-cardiac conditions} Table \ref{tab:non-cardiac} presents the predictive performance results for various non-cardiac conditions. Overall, the model demonstrates strong generalization across a diverse array of non-cardiac conditions, with internal AUROC values ranging from 0.70 to 0.95.

\section{Discussion}

\heading{Cardiac conditions}

For cardiac conditions, the use of ECG data is naturally aligned with clinical practice, as ECGs are a fundamental tool in diagnosing and managing heart-related disorders. The model's ability to accurately predict conditions like atrial fibrillation, heart failure, and conduction disorders, with AUROC values up to 0.87, underscores its potential to assist clinicians in early detection and personalized treatment planning. Accurate predictions for these conditions can lead to timely interventions, reducing the risk of complications such as stroke, sudden cardiac death, or worsening heart failure. Furthermore, the model’s strong performance in identifying more subtle conditions like valvular heart diseases and ischemic cardiomyopathy reflects its capacity to detect complex and multifactorial cardiac conditions that may otherwise be challenging to diagnose solely based on standard ECG interpretation. 

\heading{Non-cardiac conditions}
For non-cardiac conditions, the model's predictive power is particularly noteworthy given that these conditions are traditionally not associated with direct ECG interpretation. Yet, the ability to predict conditions such as polycystic ovarian syndrome, systemic lupus erythematosus, chronic kidney disease, and chronic obstructive pulmonary disease (COPD) with high accuracy (AUROC up to 0.95) suggests that ECG data may capture indirect physiological changes associated with these diseases. For instance, systemic inflammation, electrolyte imbalances, and other metabolic disturbances often manifest subtly in ECG patterns, which the model can discern. This capability could revolutionize the way non-cardiac conditions are screened and monitored, allowing for earlier detection and intervention.

\heading{Clinical significance}
The integration of AI-driven analysis of ECG data in diagnosing both cardiac and non-cardiac conditions offers significant clinical benefits. It enhances diagnostic accuracy, reduces the time needed for diagnosis, and can potentially lower healthcare costs by streamlining the diagnostic process. The model’s robust performance across a wide range of conditions highlights its versatility and potential to become an invaluable tool in diverse clinical settings, enabling screening for various conditions through ECG-based AI predictions. This approach extends beyond the cardiovascular system to include areas such as endocrine/metabolic, autoimmune/rheumatologic, musculoskeletal, ophthalmological, gastrointestinal, renal, respiratory, infectious diseases, and neurological conditions.

\heading{ECG features vs. raw signals}
The results also underscore the high information density of a relatively small set of ECG features to characterize the ECG. On the one hand, these results foreshadow additional improvements in predictive performance through the use of deep learning models applied to raw waveform data, as demonstrated in \cite{alcaraz2024mdsedmultimodaldecisionsupport,strodthoff2024prospects}. On the other hand, ECG features, unlike raw time series \cite{wagner2024explaining}, carry an immediate clinical meaning and models building on the former category are more straightforward to analyze for example through post-hoc explainability methods, as exemplified in \cite{ott2024using}.

\heading{Data and code availability} Code for dataset preprocessing and experimental replications can be found in our dedicated repository \cite{githubAI4HealthUOLCardioDiag}.

\bibliography{refs}

\begin{correspondence}
Nils Strodthoff \\
Carl von Ossietzky Universität Oldenburg
Fakultät VI - Medizin und Gesundheitswissenschaften
Department für Versorgungsforschung
Abteilung AI4Health
Ammerländer Heerstr. 114-118
26129 Oldenburg, Deutschland \\
nils.strodthoff@uol.de
\end{correspondence}

\end{document}